\documentclass[letterpaper]{jpconf}
\usepackage{graphicx}

\newcommand{\anti}[1]
{
	\overline{#1}\mbox{}
}

\begin{document}
\title{$B \to D^{(*)} \tau \nu$ at Belle}

\author{Jacek Stypula}

\address{H. Niewodniczanski Institute of Nuclear Physics, Krakow}

\begin{abstract}
Semileptonic decays of $B$ mesons to the tau lepton, despite of experimental difficulties, are of great importance. In particular they are sensitive probes of models with extended Higgs sectors. The results of studies of semitauonic $B$ decays at Belle are presented.
\end{abstract}

\section{Introduction}
$B$ decays to $\tau$ leptons represent a broad class of processes that can provide interesting tests of the Standard Model (SM) and its extensions. There is little experimental information about decays of this type due to difficulties related to multiple neutrinos in the final states. At $B$-factories $B$ decays to multi-neutrino final states can be observed via the recoil of accompanying $B$ meson ($B_{\rm tag}$). The $B_{\rm tag}$ can be reconstructed inclusively from all the particles that remain after selecting $B_{\rm sig}$ candidates or exclusively in several, mostly hadronic decay modes. Reconstruction of $B_{\rm tag}$ strongly suppresses the combinatorial and continuum backgrounds and provides kinematical constraints on the signal meson ($B_{\rm sig}$).

In this report, we present the results of studies of $B^0 \to D^{*-} \tau^+ \nu_{\tau}$ \cite{pub-dtaunu} decays.\footnote{
Charge conjugate modes are implied throughout this report unless otherwise stated.} These analysis are based on a data samples of $492$ fb$^{-1}$ recorded at the $\Upsilon(4S)$ resonance with the Belle detector \cite{detector} at the KEKB collider \cite{kekb}. It corresponds to $535 \times 10^6$ $B\anti{B}$ pairs.

\section{Analysis}
$B$ meson decays with $b \to  c \tau \nu_{\tau}$ transitions, due to the large mass of the $\tau$ lepton, are sensitive probes of models with extended Higgs sectors \cite{Itoh} and provide observables sensitive to new physics, such as polarizations, which cannot be accessed in other semileptonic decays.

The signal decay is reconstructed by selecting combinations of a $D^{*-}$ meson and a charged track expected from $\tau^+$ decay, and
the remaining particles are used to reconstruct inclusively the $B_{\rm tag}$ decay. $D^*$ mesons are reconstructed in the $D^{*-} \to \anti{D}^0 \pi^-$ decay mode. $\anti{D}^0$'s are chosen to decay to $K^+ \pi^-$ or $K^+ \pi^- \pi^0$. The $\tau$ lepton candidates are reconstructed in $\tau^+ \to e^+ \nu_e \anti{\nu}_{\tau}$ and $\tau^+ \to \pi^+ \anti{\nu}_{\tau}$ decays. We exclude $\tau^+ \to \mu^+ \nu_{\mu} \anti{\nu}_{\tau}$ due to inefficient muon identification in the relevant momentum range. 

For $\tau^+ \to \pi^+ \anti{\nu}_{\tau}$ decays only $\anti{D}^0 \to K^+ \pi^-$ mode is used to avoid the higher combinatorial background. The following variables, $M_{\rm{tag}} = \sqrt{E_{\rm beam}^2 - p_{\rm tag}^2}$ and $\Delta E_{\rm tag} = E_{\rm tag} - E_{\rm beam}$  are used to check the consistency of a $B_{\rm tag}$ with $B$ meson decay.  $E_{\rm beam}$ is the beam energy whereas $p_{\rm tag}$ and $E_{\rm tag}$ are momentum and energy of residual particles respectively. We require that events satisfy $M_{\rm tag}>5.2$ GeV/$c^2$ and $|\Delta E_{\rm tag}|<0.6$ GeV. Additional requirements such as zero total event charge, no additional leptons in the event and zero barion number, number of neutral particles on the tagging side $N_{\pi^0} + N_{\gamma}<5$ and remaining energy in ECL $E_{\rm ECL}<0.35$ GeV are imposed to improve the the $B_{\rm tag}$ purity.

In order to validate $B_{\rm tag}$ reconstruction, the control sample $B_{\rm sig} \to D^{*-} \pi^+$ is used. Monte Carlo is consistent with data. The procedure of the inclusive reconstruction selects B decays with pairs of $D^*$ and electron or $D^*$ and pion in the final state. The main sources of background are from $B \to D^* e \nu_e$ decays for $\tau^+ \to e^+ \nu_e \anti{\nu}_{\tau}$ mode and 
combinatorial background from hadronic $B$ decays for $\tau^+ \to \pi^+ \anti{\nu}_{\tau}$ channel. Background suppression employs observables that are sensitive to missing energy in $B_{\rm sig}$ decay, like missing energy $E_{\rm mis}=E_{\rm beam}-E_{D^{*}}-E_{e,\pi}$ and visible energy of the event, square of missing mass $M_M^2 = E_{\rm mis}^2 - (p_{sig}-p_{D^{*}}-p_{e,\pi})^2$ and the effective mass of the $\tau \nu_{\tau}$ system $M_W^2 = (E_{\rm beam}-E_{D^{*}})^2-(p_{\rm sig} - 
p_{D^{*}})^2$ The most effective variable $X_{\rm mis}$, is defined  by $(E_{\rm mis} - |p_{D^{*}} + p_{e,\pi}|)/\sqrt{E_{\rm beam}^2 - m^2_{B^{0}}}$ and is closely related to the missing mass in the $B_{\rm sig}$ decay.

\begin{figure}[!h]
\begin{center}
\parbox{.46\textwidth}{
\begin{center}
\includegraphics[width=.39\textwidth,height=.39\textwidth]{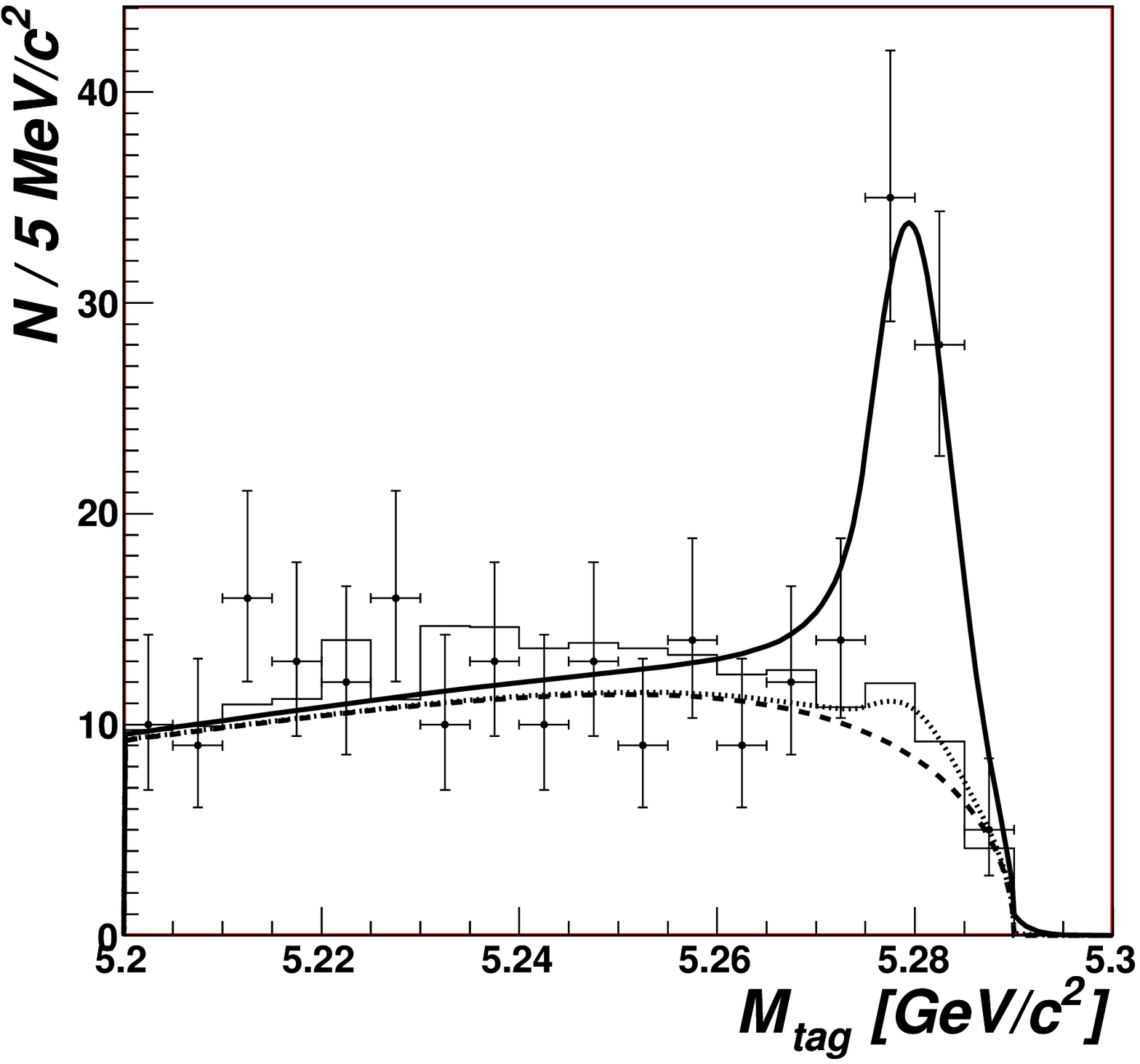}
\caption{\label{pic-dtaunu} $M_{\rm tag}$ distribution for $B^0 \to D^{*-} \tau^+ \nu_{\tau}$ decay in the data (points with error bars). 
The histogram represents the expected background.
The solid line shows the fit result. The dotted and dashed curves indicate the background 
components.}
\end{center}}\hspace{.07\textwidth}
\parbox{.46\textwidth}{
\begin{center}
\includegraphics[width=.39\textwidth,height=.39\textwidth]{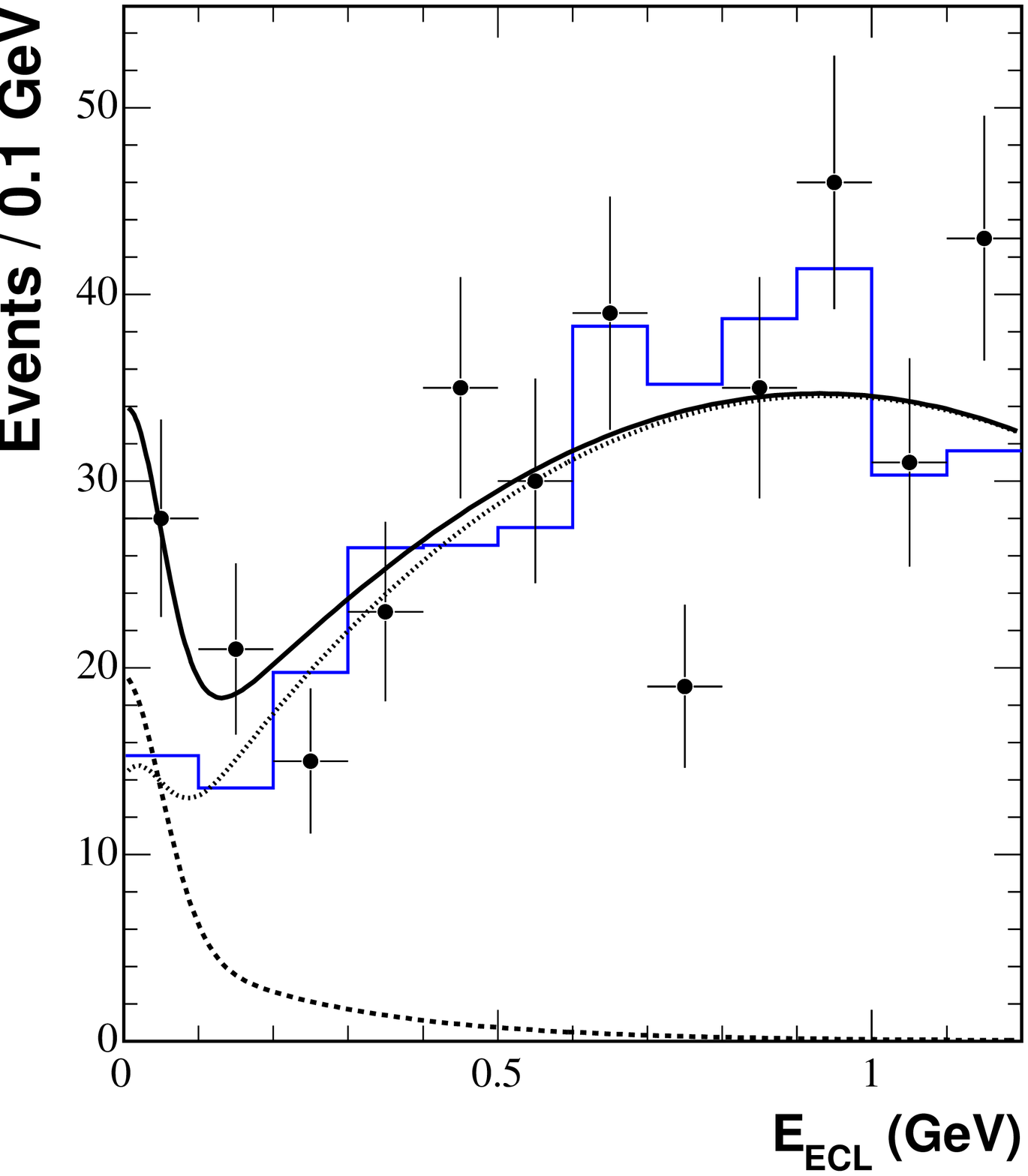}
\caption{\label{pic-taunu}$E_{\rm ECL}$ distribution for $B^+ \to \tau^+ \nu_{\tau}$ decay in the data (points with error bars).
% after all selection criteria. 
The expected background is represented by the histogram. The solid curve shows the fit result. The dashed and dotted curves represents signal and background components.}
\end{center}}
\end{center}
\end{figure}

The signal yield is extracted from an unbinned maximum likelihood fit to the $M_{\rm tag}$ distribution in the $-0.25<\Delta E_{\rm tag}<0.05$ GeV window. The result of a simultaneous fit to all analyzed subchannels constrained to a common value of $\mathcal{B} (B^0 \to D^{*-} \tau^+ \nu_{\tau})$ is shown in Figure~\ref{pic-dtaunu}. We obtain $60 ^{+12}_{-11}$ events for the $B^0 \to D^{*-} \tau^+ \nu_{\tau}$ decay. This corresponds to the branching fraction $2.02^{+0.40}_{-0.37} (\textrm{stat})\pm 0.37 (\textrm{syst})\%$, consistent with SM expectations. The significance, after including systematic uncertainties, is $5.2 \sigma$. This is the first observation of an exclusive decay with the $b \to c \tau \anti{\nu}_{\tau}$ transition.

\section{Summary}
The studies of semitauonic $B$ decays at Belle brought significant advances in this field, providing the first observation of an exclusive semi-tauonic $B$ decay in the $B^0 \to D^{*-} \tau^+ \nu_{\tau}$ channel. These results are consistent with the SM and provide valuable constraints on NP scenarios.

\end{document}